\documentclass[traditabstract]{aa}
\usepackage{graphicx}
\usepackage{natbib}
\usepackage{txfonts}
\begin{document}

   \title{The effective temperature -- radius relationship of M-dwarfs}


   \author{S.~Cassisi \inst{1,2}
          \and  M.~Salaris \inst{3}
          }

   \institute{INAF-Osservatorio Astronomico d'Abruzzo, via M. Maggini, sn.
     64100, Teramo, Italy (santi.cassisi@inaf.it)
     \and 
     INFN -  Sezione di Pisa, Largo Pontecorvo 3, 56127 Pisa, Italy
     \and Astrophysics Research Institute,
     Liverpool John Moores University, IC2, Liverpool Science Park,
     146 Brownlow Hill, Liverpool, L3 5RF, UK  
     }

   \date{Received ; accepted }

 
\abstract{M-dwarf stars provide very favourable conditions to find habitable worlds beyond our solar system.
The estimation of the fundamental parameters of the transiting exoplanets rely on the accuracy of the theoretical
predictions for radius and effective temperature of the host M-dwarf, hence the importance of multiple empirical tests
of very low-mass star (VLM) models, the theoretical counterpart of M-dwarfs. Recent determinations of 
mass, radius and effective temperature of a sample of M-dwarfs of known metallicity have disclosed a 
supposed discontinuity
in the effective temperature-radius diagram corresponding to a
stellar mass of about 0.2$M_{\odot}$, that has been ascribed to the transition from partially convective to
fully convective stars. In this paper we compare existing VLM models to these observations, and find
that theory does not predict any discontinuity at around 0.2$M_{\odot}$, rather a smooth change of slope 
of the effective temperature-radius relationship around this mass value. The appearance of a discontinuity 
is due to naively fitting the empirical data with linear segments. 
Also, its origin is unrelated to
the transition to fully convective structures. We find that this feature is instead 
an empirical signature for the transition to a regime where electron degeneracy provides
an important contribution to the stellar EOS, and constitutes an additional test of the consistency of the
theoretical framework for VLM models.}
\keywords{stars: late-type -- stars: low mass -- stars: fundamental parameters 
}

\titlerunning{M-dwarf effective temperature -- radius relationship}
\authorrunning{S. Cassisi \& M. Salaris}
   \maketitle
%

\section{Introduction}\label{intro}

M-dwarf stars may be our biggest opportunity to find habitable worlds beyond our solar system. 
This class of stars comprise $\approx$70\% of all stars in the Milky Way, and   
small earth-like planets are easier to detect when orbiting small stars via both transit and radial-velocity techniques. Also, 
the habitable zones are much closer to the host star than the case of Sun-like stars, thus increasing the probability of observing a transit
\citep[see, e.g., the review by][]{sbj}.
About 200 exoplanets have been found around M-dwarfs, many of them in their host star habitable zone \citep[see, e.g.][]{qbr,ae}. Current and
planned missions like NASA's Transiting Exoplanet Survey Satellite \citep[(TESS --)][]{ricker15}, and the ESA's PLAnetary
Transits and Oscillations of stars (PLATO) mission \citep{plato} will facilitate the discovery of several more planets hosted by M-dwarfs.

The estimation of the fundamental parameters of a transiting exoplanet, such as its mass and radius, rely on the determination 
of mass and radius of the host star, while the planet
surface temperature and the location of the habitable zone depend on the star radius and effective temperature. These determinations often involve the use very low-mass star (VLM) models --models for stars with mass in the range
between about 0.5-0.6~$M_\odot$ and the minimum mass
that ignites H-burning ($\sim 0.1M_\odot$)-- that are the theoretical counterpart of M-dwarfs \citep[see, e.g.,][for a review]{cb00}.

Comparisons of VLM calculations with empirical determinations of M-dwarf radii, masses and effective temperatures,  
are therefore crucial to assess the reliability of the models, hence the accuracy of the estimated parameters for planets hosted by M-dwarfs    
\citep[see, e.g.,][and references therein]{torres, fc, parsons}.
These empirical benchmarks, for example, have disclosed a small average offset
between theoretical and empirical mass-radius
relationships for VLM models --theoretical radii being smaller at a given mass-- by $\sim$3\% on average \citep{fc, spada, bastinew}.
This is generally ascribed to the effect of large-scale magnetic fields that are not routinely included in VLM model calculations.

Very recently, \citet{rabus19} have combined their own near-infrared long-baseline interferometric measurements obtained with the
Very Large Telescope Interferometer, with
accurate parallax determinations from \textit{Gaia} Data Release 2, to estimate mass, linear radius, effective temperature and bolometric
luminosity of a sample of M-dwarfs of known metallicity.
By implementing their own data set with the much larger sample by \citet{mann15}, they claimed to 
have found a discontinuity
in the effective temperature-radius diagram around $\sim$3200-3300~K, corresponding to a
stellar mass of about 0.23$M_{\odot}$. These authors concluded that the discontinuity is
likely due to the transition from partially convective M-dwarfs to the fully convective regime,
although no comparison with theoretical models was performed.

The most recent sets of VLM theoretical models actually predict the transition to fully convective stars 
at masses $\sim$0.35$M_{\odot}$ \citep[see, e.g.,][]{cb97, bhac, bastinew}, a value about 0.15$M_{\odot}$  larger than assumed by \citet{rabus19} to
explain the observed discontinuity.
The goal of this paper is therefore to reanalyze the results by \citet{rabus19}, by performing detailed comparisons with theoretical VLM models in the
effective temperature-radius diagram, to assess whether models predict this discontinuity and what is its actual physical origin.

In the next section we analyze the mass, radius and effective temperature data employed by these authors by 
performing detailed comparisons with theoretical VLM models, and
identify the physical reason for the discontinuity in the effective temperature-radius diagram. 
A summary and conclusions follow in Sect.~\ref{conclusions}.

\section{Analysis}\label{analysis}

The paper by \citet{rabus19} --hereafter R19-- presents empirical estimates of mass ($M$), radius ($R$) and effective temperature ($T_{\mathrm eff}$)
for 22 low-mass dwarfs with 
masses between $\sim 0.15$ and $\sim 0.55 M_{\odot}$, and [Fe/H] between $\sim-$0.6 and $\sim$ +0.5. Eighteen
of these object are in common with \citet{mann15} --hereafter M15-- who determined $M$, $R$ and $T_{\mathrm eff}$ values for
a much larger sample (over 180 objects) of the same class of objects. For the stars in common, the 
$R$ and $T_{\mathrm eff}$ values determined by
R19 are generally in good agreement (within the associated errors) with M15 (see Fig.~2 from R19).
After merging their results with the rest of M15 sample, R19 performed simple linear fits 
to the data in the $T_{\mathrm eff}$ -$R$ diagram --without using theory as a guideline-- 
finding a discontinuity of the slope at $T_{\mathrm eff}\sim$3200-3340~K, 
corresponding to a radius $R\sim 0.22~R_{\odot}$, and a mass $M\sim 0.23 M_{\odot}$.

R19 do not present any comparison with theoretical models, but explain this discontinuity 
as the signature of the transition from partially to fully convective structures.
In the following we compare theoretical VLM models with this data, demonstrating that the existence 
of a discontinuity is only apparent, due to the way the data are fit, and it is not intrinsic of 
the models. Also, the physical origin of this apparent discontinuity 
is completely different from that hypothesised by R19.

\subsection{Comparison with stellar evolution models}\label{model}

In our analysis we rely on models 
from the BaSTI-IAC database \citep{bastinew}\footnote{Models are publicly available at the following URL: http://basti-iac.oa-abruzzo.inaf.it}.
We refer to \citet{bastinew} for details about the model physics inputs and the parameter space covered by the calculations.
Here we briefly recall that the radiative Rosseland 
opacity is calculated employing the OPAL results \citep{ir:96} for temperatures larger than log($T$)=4.0, and \citet{faa05} results
--including contributions from molecules and grains-- for lower temperatures. Both high- and low-temperature opacity tables have been computed
for the solar-scaled heavy element distribution
determined by \citet{caffau}. Electron conduction opacity is calculated using the results by \citet{cassisi:07}, whilst for the equation of state (EOS) 
we employ the \lq{FreeEos}\rq\ 
by A. Irwin \citep[see][for a short discussion about the characteristics of this EOS]{csi:03}. The EOS tables have been calculated with the option
EOS1 in Irwin's code, that provides the best match to the OPAL EOS \citet{rn02}, and to the EOS  by \citet{scvh} in the low-temperature and
high-density regime relevant to VLM models. 

The temperature gradient in 
superadiabatic surface convective layers is calculated according to the mixing-length theory 
\citep{bv:58}, using the formalism by \citet{cg:68}, with the mixing-length parameter ${\rm \alpha_{ML}}$ set to 2.006 as obtained from the
standard solar model calibration \citep[see,][for more details]{bastinew}. The outer boundary conditions for the model calculations (pressure
and temperature at a Rosseland optical depth $\tau$=100) are obtained from the non-grey PHOENIX model atmosphere library
\citep{allard:12}. 

\begin{figure}
   \centering
   \includegraphics[width=8.7cm]{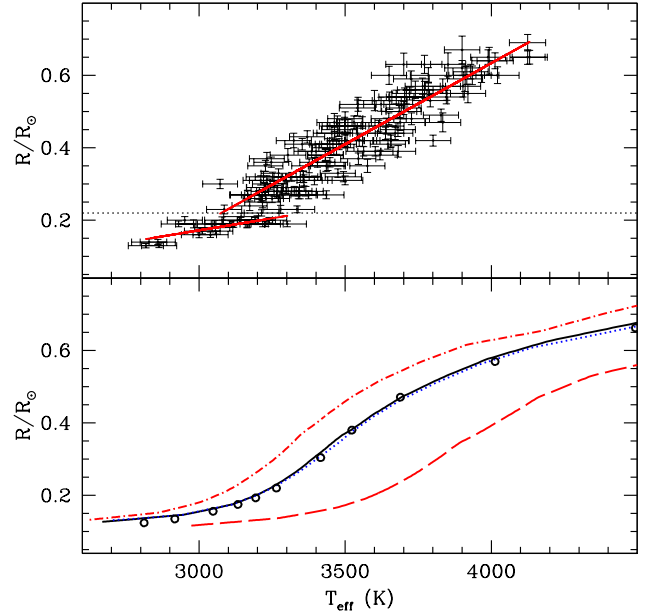}
   \caption{{\sl Upper panel}:$T_{\mathrm eff}$ -$R$ diagram for M15 data (including error bars). The solid lines show the fitted linear relationships
     discussed in the text. {\sl Lower panel}: $T_{\mathrm eff}$ -$R$ relationships from theoretical VLM models.
     Dot-dashed, solid and dashed lines display 10~Gyr BaSTI-IAC models with [Fe/H]=+0.45, +0.06 and $-$0.60, respectively.
     The dotted line shows BASTI-IAC
     1~Gyr, [Fe/H]+0.06 models, while the open circles display \citet{bhac} results for 10~Gyr and [Fe/H]=0.0.}
         \label{RTeff_obs}
   \end{figure}

The upper panel of Fig.~\ref{RTeff_obs} displays the data by M15 (with error bars) in a $T_{\mathrm eff}$ -$R$ diagram. We consider here
only the M15 sample, because these authors provide in tabular form  also [Fe/H] values in addition to $M$, $R$ and $T_{\mathrm eff}$.
Just considering M15 stars does not alter at all the conclusions by R19. 
By fitting linear relationships to this $T_{\mathrm eff}$ -$R$ diagram as in R19, the discontinuity claimed by R19 
is still clearly visible in Fig.~\ref{RTeff_obs} at
$R\sim 0.22~R_{\odot}$. A least squares linear fit to the data (also displayed in Fig.~\ref{RTeff_obs})
provides $R/R_{\odot}= 0.763(T_{\mathrm eff} /5777)-0.224$
(with a 1$\sigma$ dispersion equal to 0.01 around this mean relation) if $R/R_{\odot} <$ 0.22,
and $R/R_{\odot}= 2.583(T_{\mathrm eff} /5777)-1.155$ (with a 1$\sigma$ dispersion equal to 0.05
around this mean relation) if $R/R_{\odot} >$ 0.22 .
These values of slope and zero point are consistent with R19 results (see their Eq.~7) within their quoted error bars. Also the values of the
dispersion of the observed points around these mean linear relationships are consistent with R19 results.

The lower panel of  Fig.~\ref{RTeff_obs} displays the theoretical $T_{\mathrm eff}$ -$R$ relationships for an age of 10~Gyr, [Fe/H]=$-$0.60, +0.06 and +0.45,
masses between 0.1 and $\sim$ 0.6-0.7$M_{\odot}$, as derived from the BaSTI-IAC models.
In the same diagram we show also the $T_{\mathrm eff}$ -$R$ relationship for [Fe/H]=0.06 but an age of 1~Gyr, and the 10~Gyr, [Fe/H]=0.0
relationship derived from the independent \citet{bhac} calculations.
Theoretical models display a smooth and continuous change of slope of the $T_{\mathrm eff}$ -$R$ relationship around 
$R \sim 0.2 R_{\odot}$, at temperatures increasing with
decreasing [Fe/H]. Notice that the assumed stellar age does not affect the shape of the $T_{\mathrm eff}$ -$R$ relationship, and also that
models from different authors
give essentially the same result both qualitatively and -- at least at metallicities around solar -- quantitatively. 

\begin{figure}
     \centering
     \includegraphics[width=8.7cm]{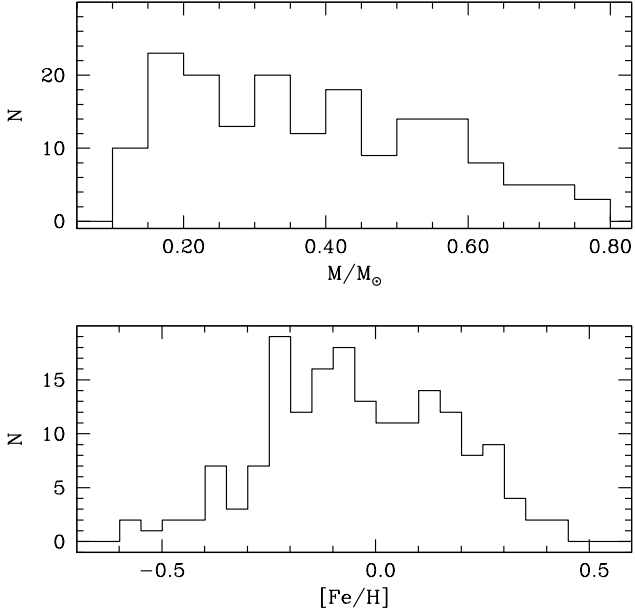}
      \caption{Mass (upper panel) and [Fe/H] (lower panel) number distribution of M15 M-dwarf sample.}
         \label{sample}
   \end{figure}

To make a close comparison with M15 data, we have created a synthetic sample of stars with our 10~Gyr BaSTI-IAC grid of models,
employing the same mass and [Fe/H] distribution of M15 sample, shown in Fig.~\ref{sample}.

\begin{figure}
   \centering
   \includegraphics[width=8.7cm]{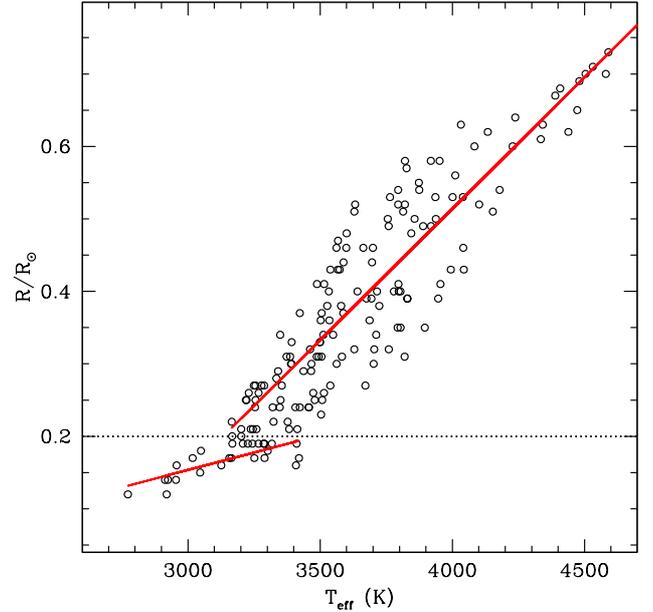}
      \caption{As the upper panel of Fig.~\ref{sample}, but for a synthetic sample built using the BASTI-IAC VLM models (see text for details).}
         \label{synth}
   \end{figure}

In brief, for each individual star in M15 sample, we have considered the mass $M$ and [Fe/H] values given by these authors,
perturbed by a Gaussian random error 
with the same 1$\sigma$ dispersions given by M15 (typically 0.08~dex for [Fe/H], and about 0.1$M$ for the mass).
With these ($M$, [Fe/H]) pairs we then interpolated amongst the model grid 
to determine the corresponding theoretical $R$ and $T_{\mathrm eff}$ values.
We have repeated this procedure several times to create 100 synthetic counterparts of M15 sample.
In each case a clear change of slope in the $T_{\mathrm eff}$ -$R$ relation does appear. Figure~\ref{synth} shows one of our
synthetic samples, that is representative of the overall result.

A least squares linear fit to the data provides $R/R_{\odot}= 0.552(T_{\mathrm eff} /5777)-0.133$
(with a 1$\sigma$ dispersion equal to 0.01 around this mean relation) if $R/R_{\odot} <$ 0.20 ,
and $R/R_{\odot}= 2.094*(T_{\mathrm eff} /5777)-0.936$ (with a 1$\sigma$ dispersion equal to 0.06 
around this mean relation) if $R/R_{\odot} >$ 0.20. These two linear relationships are also shown in Fig.~\ref{synth}.
The discontinuity claimed by R19 is retrieved in this synthetic sample, despite the lack of discontinuities in 
the theoretical models, and it is due to fitting the data in this diagram with linear segments. It simply reflects the smooth 
change of slope of the model $T_{\mathrm eff}$ -$R$ relation.

\begin{figure}
   \centering
   \includegraphics[width=8.7cm]{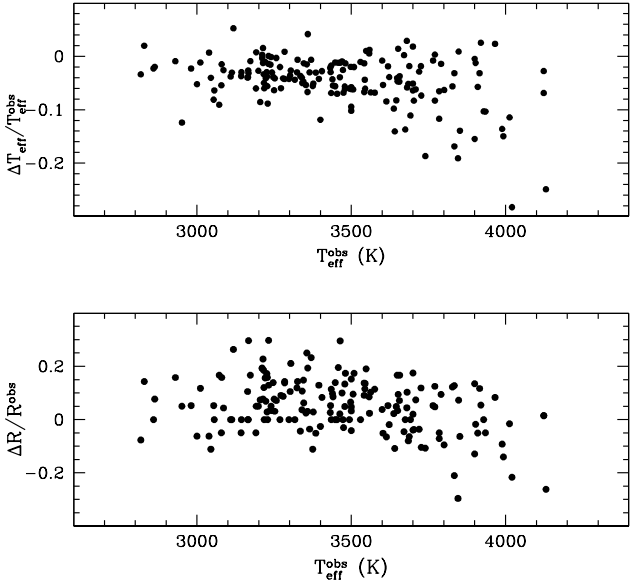}
   \caption{{\sl Upper panel}: Fractional difference (observations-theory) of $T_{\mathrm eff}$ as a function of the empirical temperatures
     ($T_{\mathrm eff}^{obs}$),
     between M15 data and the synthetic sample of Fig.~\ref{synth}.
     {\sl Lower panel}: As the upper panel, but for the difference between empirical ($R^{obs}$) and theoretical radii (see text for details).}
         \label{diff}
   \end{figure}

The synthetic samples share qualitatively the same properties of the observed one, although quantitatively there are some differences. The change
of slope in the $T_{\mathrm eff}$ -$R$ relation appears at slightly lower radii compared to M15. Also, the actual values of slopes and zero points are
slightly different from the empirical result.
The reason for these differences becomes clear when examining Fig.~\ref{diff}, that displays fractional differences of 
$R$ and  $T_{\mathrm eff}$ between the synthetic sample of Fig.~\ref{synth} and M15 --calculated as (observations-theory)--
as a function of M15 $T_{\mathrm eff}$ estimate.

For observed $T_{\mathrm eff}$ values up to $\sim$3600~K (corresponding to stellar masses $\sim 0.45 M_{\odot}$)
there are systematic average constant offsets between theory and observations, and no trends with
the empirical $T_{\mathrm eff}$. In this range the average difference in radius is 8$\pm$9 \% (observed radii being larger) 
and 3$\pm$3 \% in $T_{\mathrm eff}$ (model temperatures being larger).
At higher temperatures there are trends with the observed $T_{\mathrm eff}$, in the direction of both model temperatures and radii becoming increasingly larger
compared to observations, and an increasing spread of the differences at a given $T_{\mathrm eff}$ .

Varying the age assumed for the stars in M15 sample from 10~Gyr to 1~Gyr does not alter significantly --the change of the model
$R$ and $T_{\mathrm eff}$ for masses up to $\sim$0.4$M_{\odot}$ is almost zero-- the results for both the slopes in the $T_{\mathrm eff}$ -$R$ diagram,
and the differences of $R$ and $T_{\mathrm eff}$ between models and observations, because of the very slow evolution of stars in this mass range.

\subsection{Why a change of slope of the $T_{\mathrm eff}$ -$R$ relationship?}\label{theo}

\begin{figure}
   \centering
   \includegraphics[width=8.7cm]{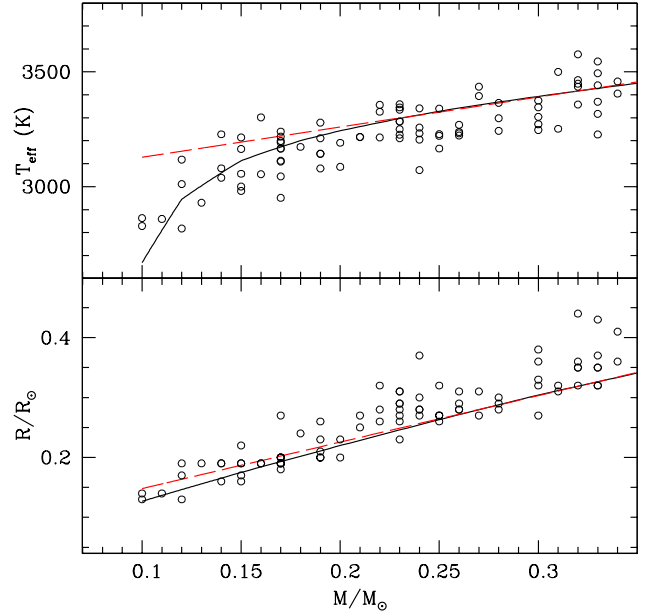}
   \caption{{\sl Upper panel}: $M$-$T_{\mathrm eff}$ relation around M=0.2$M_{\odot}$, for the M15 sample (open circles) and
     the 10~Gyr BaSTI-IAC models with [Fe/H]=0.06. {\sl Lower panel}: As the upper panel, but for the $M$-$R$ relation.
     Dashed red lines in both panels denote linear fits to the model results in the mass 
     range between 0.35 and 0.25 $M_{\odot}$ (see text for details).}
         \label{trm}
    \end{figure}

Figure~\ref{trm} displays the theoretical $M$-$R$ and $M$-$T_{\mathrm eff}$ diagrams in the VLM regime, for an age of
10~Gyr (again, the choice of the age is not critical) and a representative metallicity [Fe/H]=0.06. The empirical data are also shown.  
Around 0.2$M_{\odot}$  both theoretical and empirical $M$-$T_{\mathrm eff}$ relationships
display a clear slope change, the effective temperature starting to decrease with mass faster than at higher masses. Figure~\ref{trm}
gives a clearer visual impression of this effect by showing also a comparison
between the results of a linear fit to the theoretical relationship in the mass range between 0.4 and 0.25 $M_{\odot}$, and the actual calculations.

The same holds true for the radius, even though the effect is not obvious in the data. Analogous to the case of the $M$-$T_{\mathrm eff}$ relationship,
we display in Fig.~\ref{trm} also a comparison
between the linear fit to the theoretical $M$-$R$ relation in the mass range between 0.4 and 0.25 $M_{\odot}$, and the actual results.
Clearly, also the theoretical values for the radius start to decrease faster with mass when $M$ decreases below $\sim$0.2$M_{\odot}$.
This same behaviour of radius and effective temperature with mass is predicted by \citet{bhac} models \citep[see also, e.g.,][]{bhl, bc96, cb00}.
The combination of these two slope changes causes the change of slope of the $T_{\mathrm eff}$ -$R$ relationship.

The physical origin of this phenomenon is certainly not the transition to fully convective stars, that happens at larger masses.
The culprit is the electron degeneracy, as shown by \citet{cb00} and with more details by \citet{cb97} --see e.g. Figs. 6, 12 and 13 
in \citet{cb97}. For masses below $\sim$0.2~$M_{\odot}$ electron degeneracy 
starts to provide a sizable contribution to the gas EOS. Therefore, at around 0.2~$M_{\odot}$ the model $M$-$R$ relation begins to change slope (faster
decrease with $M$) to eventually approach the $M \propto R^{-3}$ relation below the H-burning limit. As an example, when extrapolated at 0.1~$M_{\odot}$, the
the $M$-$R$ linear relation in the regime between 0.4 and 0.25~$M_{\odot}$ (see Fig.~\ref{trm}) provides a radius equal to $\sim$0.15$R_{\odot}$.
The actual calculations give $R \sim 0.125 R_{\odot}$, whilst the zero-temperature degenerate $M$-$R$ relation for
solar chemical composition provides $R\sim 0.07 R_{\odot}$.

\begin{figure}
   \centering
   \includegraphics[width=8.7cm]{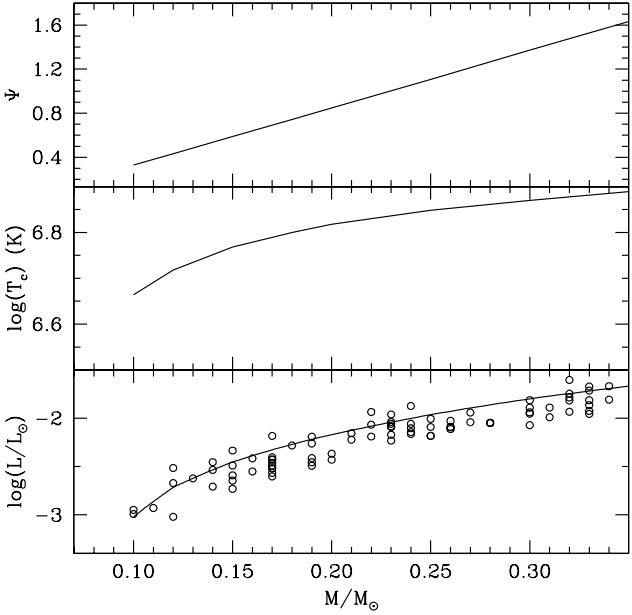}
   \caption{{\sl Upper panel}: Theoretical relationship between central degeneracy parameter and total stellar mass around M=0.2$M_{\odot}$, 
     for 10~Gyr BaSTI-IAC models with [Fe/H]=0.06.. 
     {\sl Middle panel}: As the upper panel, but for the central temperature. 
     {\sl Lower panel}: As the upper panel but for the surface luminosity. Data from M15 is also displayed (see text for details).}
         \label{figtcrc}
   \end{figure}

Figure~\ref{figtcrc} displays the relationships between central temperature ($T_c$) and $M$, and central degeneracy parameter
($\psi=k_B T/E_F$ where $k_B$ is the Boltzmann constant and $E_F$ the electron Fermi energy) and $M$, in the same mass range
as in Fig.~\ref{trm}. It is very clear the steady decrease of $\psi$ and $T_c$ with decreasing mass. The rate of decrease of $T_c$
also changes around 0.2~$M_{\odot}$.
Around this mass, the ever increasing contribution of the electron degeneracy pressure accelerates the decrease of $T_c$ with $M$, that in turn
causes a faster reduction of the efficiency of the H-burning via the $p-p$ chain. As a consequence, also the surface luminosity decreases faster with
decreasing mass below 0.2~$M_{\odot}$.
The trend of the luminosity with $M$  is also shown in Fig.~\ref{figtcrc}, together with the empirical data by M15,
that follows the trend predicted by the models.

The change of slope of the $M$-$T_{\mathrm eff}$ relationship in Fig.~\ref{trm} is then a consequence of the change of the trends of $L$ and $R$ with
mass at  M=0.2$M_{\odot}$, 
given that $T_{\mathrm eff}=(L/(4 \pi \sigma R^2))^{1/4}$ \citep[where $\sigma$ is the Stefan-Boltzmann constant, see also][]{cb97}.

\section{Summary and conclusions}
\label{conclusions}

Recent empirical determinations of mass, effective temperature and radius for a large sample of M-dwarfs (M15, R19) have disclosed
a supposed discontinuity in the $T_{\mathrm eff}$ -$R$ diagram, corresponding to a mass $\sim$0.2$M_{\odot}$. R19 have hypothesised
that the reason for this discontinuity is the transition to fully convective stars, although they did not perform any comparison with theory.

Here we have compared a set of existing theoretical VLM models to these observations, to assess whether this 
discontinuity is predicted by theory, and its physical origin. 
Theoretical models do not show any discontinuity in the $T_{\mathrm eff}$ -$R$ diagram, rather a smooth change of slope 
around $\sim$0.2$M_{\odot}$. The discontinuity found by R19 arises from fitting the empirical data with linear segments.
To this purpose, we have created synthetic samples of stars with the same [Fe/H], mass and error distributions of the observations, 
fitting the resulting $T_{\mathrm eff}$ -$R$ diagram with linear segments, as done by R19.
Despite some small offsets between the $T_{\mathrm eff}$  and $R$ scale of the models and the observations, the linear fits 
show a discontinuity corresponding to a mass $\sim$0.2$M_{\odot}$, where the theoretical models display a smooth change of slope. 
As discussed by \citet{cb97}, this smooth change of slope of the models reflects the growing
contribution of electron degeneracy to the gas EOS below $\sim$0.2$M_{\odot}$, 
and not the transition to fully convective stars, that is predicted to happen at
$\sim$0.35$M_{\odot}$. A stronger electron degeneracy causes a faster decrease of $R$ with decreasing $M$, and
a faster decrease of both central temperature and luminosity with decreasing $M$. These, in turn, 
induce a steeper decrease of $T_{\mathrm eff}$ with $M$.
These changes of slope of the $M$-$R$ and $M$-$T_{\mathrm eff}$ produce a corresponding change in 
the $T_{\mathrm eff}$ -$R$ diagram.
The empirical results by M15 and R19 provide therefore a clear signature of the threshold beyond which electron degeneracy provides 
an important contribution to the stellar EOS. This data are therefore an additional 
test of the consistency of the theoretical framework for VLM models.   
   
\begin{acknowledgements}
SC acknowledges support from Premiale INAF MITiC, from INFN (Iniziativa specifica TAsP),  PLATO ASI-INAF contract n.2015-019-R0 \& n.2015-019-R.1-2018, and 
grant AYA2013-42781P from the Ministry of Economy and Competitiveness of Spain.
The authors acknowledge the anonymous referee for pointing out the discussion in Sect. 3.1.1. of \citet{cb97}.
\end{acknowledgements}

\bibliographystyle{aa}
\bibliography{paper_cassisi}

\end{document}